\documentclass[%
aps,
reprint,
superscriptaddress,
 amsmath,amssymb,
prper,
floatfix
]{revtex4-2}

\usepackage{multirow}
\usepackage{graphicx}
\usepackage{dcolumn}
\usepackage{bm}
\usepackage{tikz}
\usepackage{pgfplots}
\usepackage[export]{adjustbox}
\pgfplotsset{compat=1.17}
\usepackage{listings}
\usepackage{array}
\usepackage{booktabs}

\begin{document}

\title{Stress Testing a Multimodal Large Language Model Grading Assistant on\\ Kinematics Graph Reading}

\author{Ons Zammel}
\email{ozammel@fe.up.pt}
\affiliation{LIACC - Laboratory of Artificial Intelligence and Computer Science, Faculty of Engineering, University of Porto, Porto, Portugal
}

\author{Carlos Soares}
\email{csoares@fe.up.pt}
\altaffiliation[also at ]{Fraunhofer Portugal AICOS, Porto, Portugal and BrightFactory, Portugal}
\affiliation{LIACC - Laboratory of Artificial Intelligence and Computer Science, Faculty of Engineering, University of Porto, Porto, Portugal}

\author{Fl\'avia Carvalhido}
\email{fcarvalhido@fe.up.pt}
\affiliation{LIACC - Laboratory of Artificial Intelligence and Computer Science, Faculty of Engineering, University of Porto, Porto, Portugal
 }
 
\author{Gerd Kortemeyer}
\email{kgerd@ethz.ch}
\altaffiliation[also at ]{Michigan State University, East Lansing, MI 48823, USA}
 \affiliation{%
Rectorate and ETH AI Center, ETH Zurich, 8092 Zurich, Switzerland
}%

\date{\today}

\begin{abstract}
Multimodal large language models (MLLMs) can recognize graph
features yet award credit without sufficient graphical support.
This exploratory study examines selected grading outputs for
synthetic responses to four kinematics exercises, generated
through visual distortion and progressive transformation toward
reference graphs. Different model and prompt configurations were
used. Some judgments correctly distinguished instantaneous zero
velocity from rest over a finite interval and recognized
positive curvature. Yet a severely distorted velocity--time
graph received full credit despite disconnected and branching
traces and an unclear zero-velocity reference. Other submissions
graded under the same configuration lost substantial credit.
Numerical cutoffs intended to distinguish meaningful features
from drawing irregularities were supplied without a coordinate
convention for unnumbered graphs, leaving the basis of exact
partial-credit awards unresolved. Stress testing can expose
unsupported awards before classroom use. For the configurations
examined, we do not recommend accepting scores as final grades
without instructor review, including review of full-credit
responses.
\end{abstract}

\maketitle

\section{Introduction}

Multimodal large language models (MLLMs) have been successfully used to judge assessments that allow for partially correct, open-ended submissions against a so-called ``rubric'' or ``grading scheme''~\cite{Kortemeyer_2024_Thermodynamics,Kortemeyer_Noehl_2025}, including short-answer exams~\cite{yan2024exploring,kortemeyer2024performance}. The rubric typically spells out which statements, derivations, conclusions, or results are rewarded with points. Unlike multiple-choice or exact-match grading, rubrics require the grader to reason about the \emph{degree} of correctness~\cite{fussell2026open}; unfortunately, it is precisely this reasoning that is hardest to validate.

For an instructor, a useful graphical grader must tolerate
drawing imperfections while distinguishing responses that
differ in the physics required by the exercise. Correctly
grading a clean example does not establish how the grader
will handle a missing stopping interval, an ambiguous zero
level, or a drawing that only partly expresses acceleration.
Stress testing provides deliberately varied inputs for
examining such cases.

We examine selected grading outputs for synthetic responses
to four kinematics exercises, generated by distorting reference
graphs and transforming graphs from other exercises toward
reference solutions. Our aim is to identify what graphical
evidence supports the criterion-level awards and where the
rubric or its application leaves the judgment unresolved.
The cases inform the review of exercise instructions,
rubrics, and grading prompts before their use with student
work.

\subsection{Research questions}
\label{sec:research-questions}

The analysis of selected grading outputs is organized around
three research questions:

\begin{enumerate}
    \item \label{rq:physical-interpretation}
    \textbf{Physical interpretation.}
    How do the recorded awards and explanations reflect
    distinctions between graph values, slopes, and curvature,
    and between an instantaneous condition and a condition
    sustained over a finite interval?

    \item \label{rq:graphical-evidence}
    \textbf{Evidence supporting credit.}
    What graphical evidence supports or challenges the
    awarded credit when a submission is visually degraded,
    fragmented, or otherwise ambiguous?

    \item \label{rq:numerical-criteria}
    \textbf{Numerical criteria and partial credit.}
    How are numerical cutoffs intended to implement
    qualitative physics requirements interpreted in the
    reported explanations, and what basis do those
    explanations provide for awarding partial credit?
\end{enumerate}

We address these questions through selected cases, comparing
the submitted images and exercise requirements with the
criterion-level awards and reported explanations. The focus
is on whether the physical information available to the
grader supports its judgments.
\section{Background}
Interpreting graphs in physics requires connecting visual features to the quantities and physical situations they represent~\cite{Beichner_1994}. Research on kinematics has documented confusion between graph height and slope, difficulties interpreting areas under curves, and the treatment of graphs as pictures of an object's path~\cite{McDermott_1987,Beichner_1994}. Analyses of students' explanations have also documented
context-dependent graph interpretation strategies and confusion
between information at a point and over an
interval~\cite{Ivanjek_2016}. These difficulties extend beyond mathematical graph-reading skills: students' approaches to linear kinematics graphs varied with context, and qualitative understanding or context-free slope-calculation skills did not necessarily translate into correct determinations of instantaneous speed~\cite{Bollen_2016}. Difficulties also occur without a physics context: some students
in a multivariable calculus course gave answers consistent with
a function's second derivative when asked about its first
derivative~\cite{Christensen_2012}. Assessing a graphical response therefore requires attention to the physical meaning of its axes and features.

The revised Test of Understanding Graphs in Kinematics (TUG-K)
distinguishes assessment objectives such as determining
velocity from a position--time graph, determining acceleration
from a velocity--time graph, and relating graphs to verbal
descriptions of motion~\cite{Zavala_2017}.

Automated assessment of kinematics graphs predates the use of multimodal language models. Function plot response enabled students to construct curves that were evaluated against instructor-defined mathematical conditions, including conditions on slope and curvature~\cite{Laverty_2012}. These conditions can admit multiple correct curves without requiring a match to a single reference graph. The system evaluated the mathematical representation of the submitted curve directly. Grading graphs supplied as images adds the task of extracting the relevant features from their visual representation. The present study examines how an MLLM interprets these features and applies grading criteria to synthetic graph images.

MLLMs can describe an appropriate solution strategy while
misinterpreting the graphical information needed to apply
it, as observed in ChatGPT-4 responses to the
TUG-K~\cite{Polverini_2024_TUGK}. Multilingual concept-inventory
testing, including the TUG-K, also found weaker performance
on items requiring image interpretation than on textual
items~\cite{Kortemeyer_2025_Multimodal}. Other comparisons
report continuing difficulties with visual interpretation and
differences among models~\cite{Polverini_2024_Chatbots,
Polverini_2025_BEMA,Polverini_2025_VisualTasks}.
These studies concern models answering physics questions;
grading student-generated graphs requires separate evaluation.

Research on grading has demonstrated both useful capabilities and dependencies on task and workflow. An early feasibility study combined handwriting transcription with GPT-4 to assess synthetic handwritten physics solutions~\cite{Kortemeyer_2023_Grading}. In a study of a handwritten thermodynamics examination, transcription and rubric granularity affected scoring, while hand-drawn process diagrams were graded less reliably than mathematical derivations~\cite{Kortemeyer_2024_Thermodynamics}. For typed explanations of physics reasoning, GPT-4o reached 70--80\% agreement with human graders after rubric clarification and aggregation of five scoring runs; this was comparable to agreement between human graders on the studied problems~\cite{Chen_Wan_2025}.

Direct image-based grading also shows variation across response types. In a handwritten general chemistry examination, providing exam and rubric pages as images yielded stronger agreement with human scores for textual and chemical-reaction responses than for numerical and graphical tasks~\cite{Cvengros_2026}. An exploratory study of feedback on handwritten kinematics and dynamics solutions combined student perceptions with expert evaluations: students generally valued the feedback's clarity and usefulness, while trust and perceived technical accuracy were more variable~\cite{Tipaz_2026}. At the system level, assistants such as Ethel incorporate course materials, problem statements, reference solutions, and rubrics into the context supplied to a language model~\cite{Kortemeyer_2024}. These developments make graphical grading and feedback practically accessible while preserving the need to evaluate the resulting judgments against the intended assessment criteria.

Similarity measures require particular care in such evaluations. For transcribed handwritten physics solutions, embedding-based similarity was only modestly associated with score similarity, and responses grouped into the same cluster could receive different scores~\cite{Tauschhuber_2026}. Complementary work in computer vision has used synthetic ambiguous inputs, clustering, and attribution methods to inspect representative cases near the decision boundaries of binary image classifiers~\cite{Gomes_2024}. This provides a methodological precedent for diagnostic stress testing, although applying it to physics grading requires distinguishing changes in visual presentation from changes in the physical criteria satisfied. The present study examines this relationship through synthetic perturbations and reverse-generation sequences of kinematics graphs, using criterion-level awards and reported explanations to identify cases for closer inspection.

\section{Methods}
\label{sec:methods}

\subsection{Study design}
\label{sec:study-design}
\begin{figure}
\centering
\includegraphics[width=\linewidth]{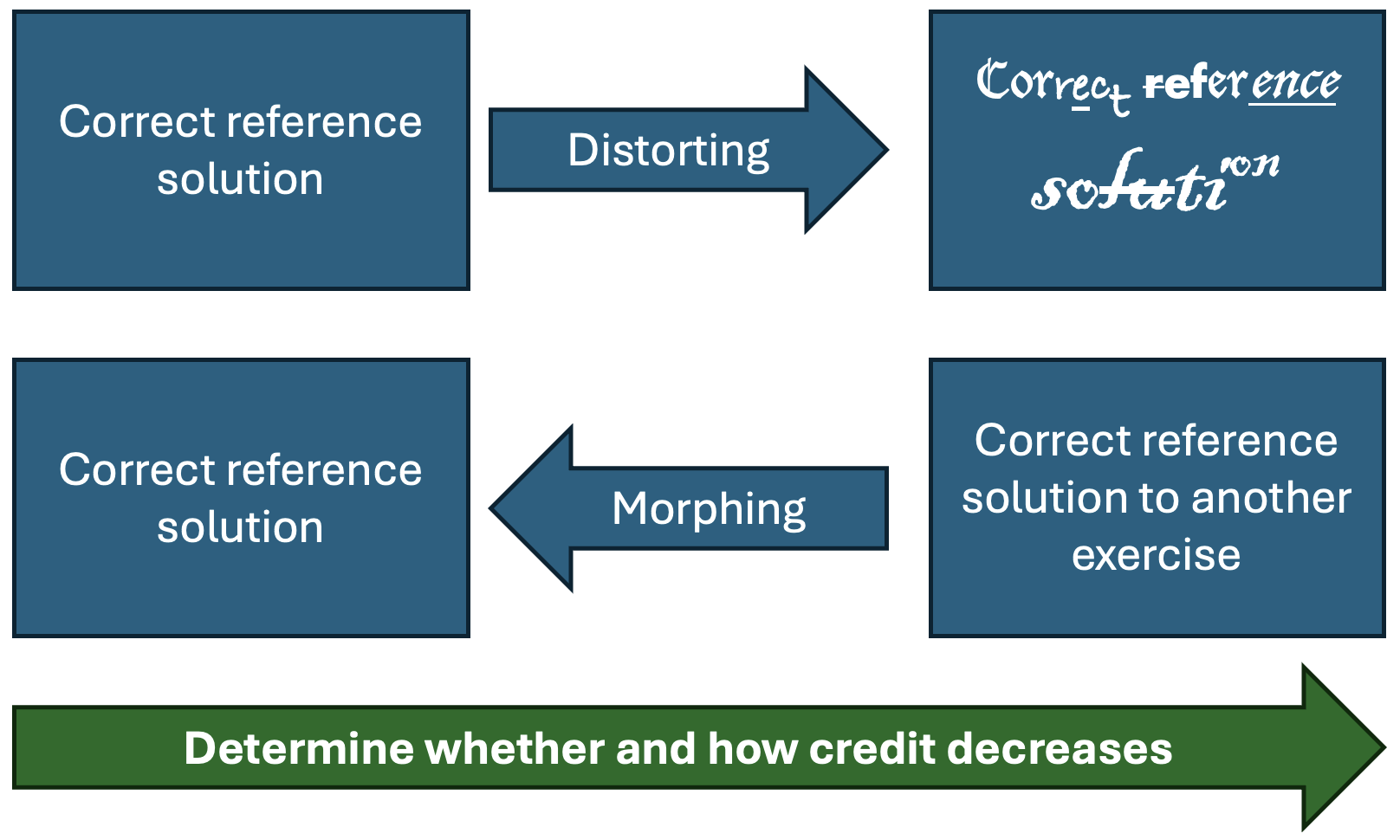}
\caption{Two procedures for generating inputs for grading.
Forward generation introduces visual alterations into a
reference graph for the target exercise. Reverse generation
progressively transforms a donor graph from another exercise
toward the target reference. The blue arrows indicate the generation directions. The green
arrow indicates the investigation of whether and how awarded
credit decreases away from the target reference; it does not
represent an established ordering of physical correctness
or scores.}
\label{fig:forward_backward}
\end{figure}

The study uses four qualitative kinematics scenarios based on
earlier work with our deterministic function plot response system~\cite{Laverty_2012} (Sec.~\ref{sec:exercises}). We generated synthetic responses
through visual perturbation (``distortion'') of reference
graphs and progressive transformation (``morphing'') of
graphs from other exercises toward a target reference
(Fig.~\ref{fig:forward_backward}). These responses were used
to examine how an MLLM-based grading assistant applies the
exercise-specific criteria.

For selected examples, we compare the submitted image and accompanying assessment information with the grader’s criterion-level awards and explanations. We examine which physical features support the recorded judgments and where the available information leaves a criterion ambiguous or underdetermined. The explanations are treated as statements to be checked against the inputs, rather than as direct evidence of the model’s internal reasoning.

Given the limited number of exercises ($n_{\mathrm{exercise}} = 4$), this work is considered an exploratory study whose results may help identify initial patterns and guide future research on the pedagogical use of MLLM graders.

\paragraph*{Problem formulation.}
Both generation procedures use the same assignment representation.
An assignment consists of a problem statement $s$, a reference graph $a$,
and a rubric $r$ containing $m$ separately scored criteria:
\begin{equation}
q = (s, a, r).
\label{eq:assignment}
\end{equation}
The reference graph illustrates one possible answer. Other graphs may satisfy the same physical criteria.

Criterion $k$ has a maximum of $w_k$ points. The maximum score for the
assignment is
\begin{equation}
W = \sum_{k=1}^{m} w_k.
\end{equation}

For a submitted graph image $x$ and a given grading output, let
$\hat{\varphi}_k(x) \in [0,1]$ denote the fraction of the available
points awarded by the grader for criterion $k$. A value of 0 denotes
no credit, and a value of 1 denotes full credit. These fractions
describe the recorded awards. Whether those awards are justified
by the physical features of the submitted graph is examined separately.

Here, $x$ denotes the submitted image; $x(t)$ in the exercise
descriptions denotes physical position.

\paragraph*{Grading function.}
The grader receives the submitted image $x$, the problem statement $s$ and
the rubric $r$. The total awarded score is
\begin{equation}
G(x,s,r)=\sum_{k=1}^{m} w_k\,\hat{\varphi}_k(x).
\label{eq:grading}
\end{equation}

\paragraph*{The score ratio.}
For a fixed assignment and grading configuration, the normalized
score is
\begin{equation}
\rho_G(x) = \frac{G(x, s, r)}{W}.
\label{eq:ratio}
\end{equation}

\subsubsection{Forward direction: visual perturbations of reference graphs}
\label{sec:forward}

For an assignment $q_i = (s_i, a_i, r_i)$, the forward
procedure starts from the reference graph $a_i$ and generates
visual variations using Stable Diffusion v1.5 image-to-image.
The accompanying text prompt is constructed programmatically
from a common sketch-style description and phrases derived
from the exercise rubric. An example is shown in
Listing~\ref{lst:img_forward} in Appendix~\ref{app:img_prompt}.
Generation strength and noise parameters were searched by
Bayesian optimization and by Optuna's tree-structured Parzen
estimator~\cite{10.1007/978-3-032-41086-3_13}.
The forward corpus comprised 300, 300, 296, and 300 graded
submissions for Exercises 1--4~\cite{10.1007/978-3-032-41086-3_13}.
Each generated submission is graded with $G$. These variations
provide inputs for inspecting changes in the grader's
criterion-level awards and explanations. We use stress testing
in the sense of examining a model with deliberately challenging
or unusual inputs~\cite{Gomes_2024}.

A visual alteration may preserve or change the physical features
required by the exercise. Visual difference from the reference
therefore does not by itself establish that a submission is
incorrect. For selected examples, we examine whether the awarded
credit is supported by the submitted graph and the target rubric.

\subsubsection{Reverse direction: progressive transformation of donor graphs}
\label{sec:reverse generation}

For a target assignment $q_i = (s_i, a_i, r_i)$, reverse generation
starts from a donor graph $\hat{a}$ taken from a different
assignment $q_j = (s_j, a_j, r_j)$. The donor may already satisfy
some criteria of the target rubric; its score under that rubric
is therefore not assumed to be zero.

A reverse trajectory consists of the donor graph followed by
successive transformed images directed toward the target reference:
\begin{equation}
X_i = \left(x^{(0)}, \ldots, x^{(K)}\right),
\qquad x^{(0)} = \hat{a},
\label{eq:trajectory}
\end{equation}
where $K=5$ is the number of transformation steps used in this
study. The generated images $x^{(1)},\ldots,x^{(5)}$ were graded against the
target assignment. They were produced without a learned model: the curve
was extracted from the donor and reference images and blended toward the
reference curve by a fraction $(k/5)^{e}$ at step $k$, where the exponent
$e \in [0.45, 1.90]$ was set per trajectory from parameters tuned with
Optuna. The reference graph $a_i$ was neither supplied to the grader nor graded.
For each exercise, 300 trajectories (3 rounds $\times$ 100) were generated,
giving 1{,}500 graded images per exercise and 6{,}000 in total. The final generated image is not assumed to be identical to $a_i$. The step number
records the order of generation; it does not determine which physical criteria the image satisfies. Where axis labels are absent or unreadable, the target exercise
specifies the intended physical quantities; any zero reference
or numerical scale inferred by the grader beyond the supplied
information is treated as an interpretive assumption when
examining the recorded grading output.

\subsection{Instruments}
\label{sec:instruments}

\subsubsection{Physics exercises}
\label{sec:exercises}
We used the following four exercise scenarios concerning graphical representations of one-dimensional kinematics:

\begin{enumerate}
    \item Provide a sketch of \(x(t)\) that always increases.

    \item A car initially goes forward, then stops for a few seconds, and then goes backward. Draw position vs time.

    \item A car initially accelerates forward from rest for a few seconds, then drives with constant speed. Draw position vs time.

    \item A car initially goes forward, then stops for a few seconds, and then goes backward. Draw velocity vs time.
\end{enumerate}

Each exercise is accompanied by an instruction, a rubric $r$,
and a reference graph $a$.

The verbal scenarios specify qualitative motion. The implemented
rubrics encode these requirements through comparisons involving
the plotted value or its derivatives. Numerical cutoffs such as
$0.05$ and $\pm0.1$ were intended as operational margins for
distinguishing clearly expressed graphical features from minor
drawing irregularities. They were not intended to impose
additional minimum physical speeds or accelerations beyond the
qualitative exercise requirements. The grader received these
numerical values without attached units or an explanation of
this intended role. The grading instructions also did not
specify a coordinate convention for applying the cutoffs to
graphs with unnumbered axes.

Once the plotted quantities are identified, features such as
increasing position, a sustained horizontal position--time
interval, and positive curvature can be assessed without
numerical axis values. An operational cutoff for detecting
such features can also be applied without physical tick
values if it is defined in plotting coordinates, for example
coordinates normalized to the graph's horizontal and vertical
ranges. Its numerical meaning nevertheless depends on that
coordinate convention. We therefore distinguish the qualitative
physics requirement, its numerical implementation, and the
points awarded. The exercise instruction specifies the order
of the motion phases and the required intervals.

\begin{figure}
\centering
\includegraphics[width=\columnwidth]{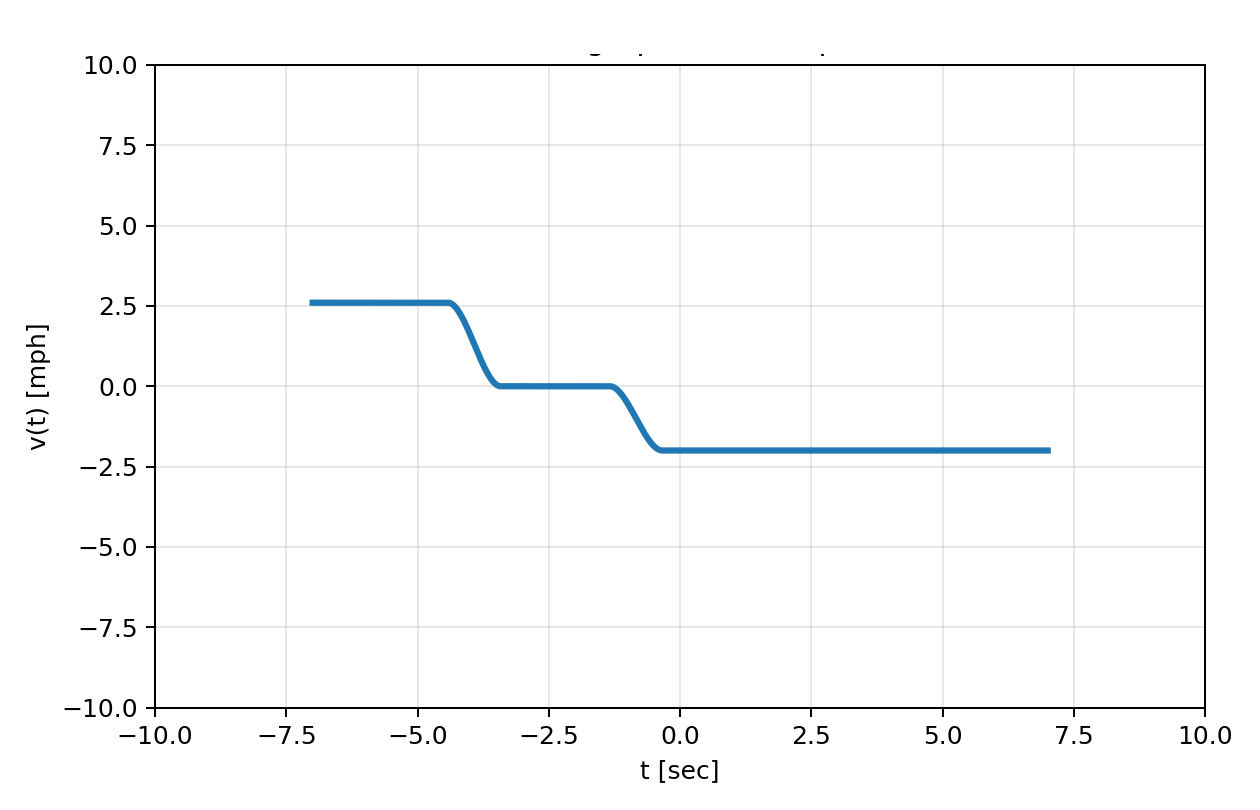}
\caption{One reference graph for Exercise 4. Positive, zero, and negative velocity values represent forward motion, rest, and backward motion, respectively.}
\label{fig:ex4}
\end{figure}

The original graph configurations use seconds for time, meters
for position in Exercises~1--3, and miles per hour (mph) for
velocity in Exercise~4. The Exercise~4 reference graph
(Fig.~\ref{fig:ex4}) is labeled in mph. Where numerical axis labels
are present, the reverse-generated Exercise~4 graphs use
$\mathrm{m/s}$. The reference graph was used in generation but
was not supplied to the standalone grader.

\begin{figure}
\centering
\includegraphics[width=\columnwidth]{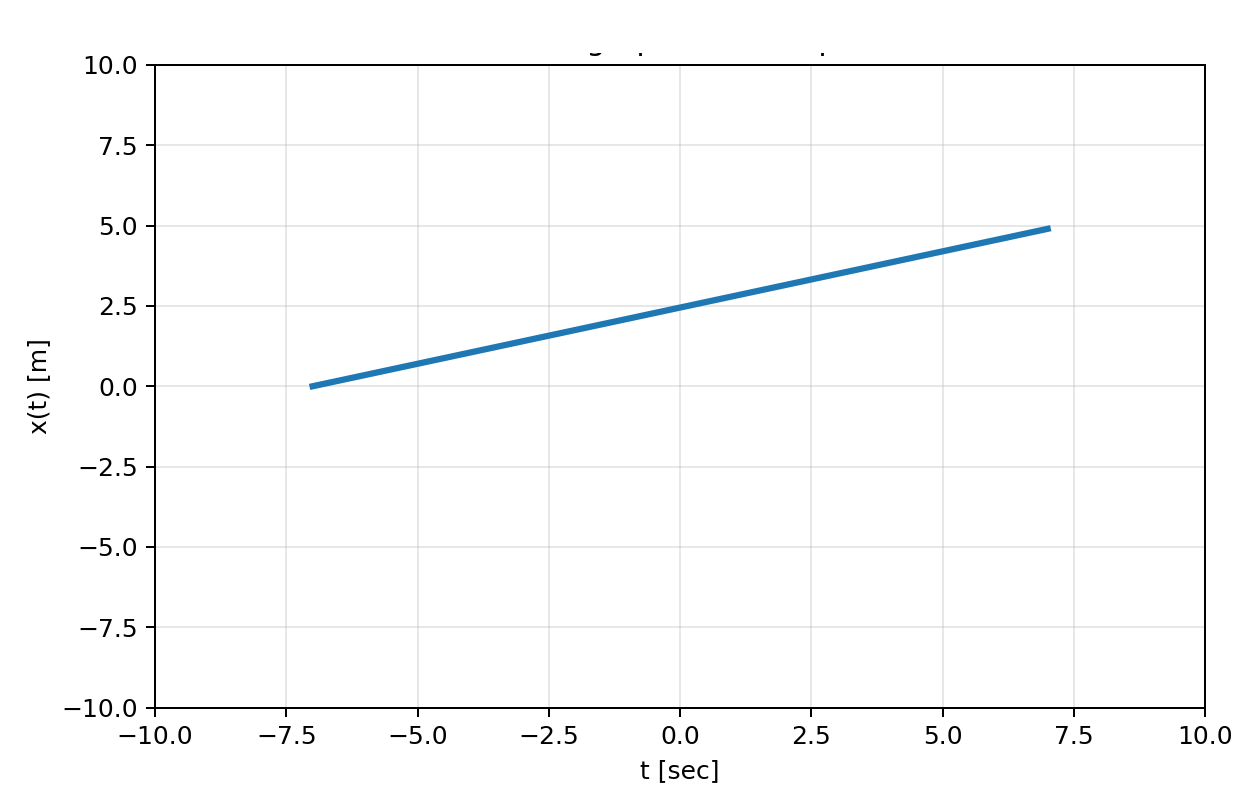}
\caption{One reference graph for Exercise 1, showing a constant positive slope.}
\label{fig:ex1}
\end{figure}

\paragraph*{Ex.\,1, Increasing position.}
Exercise 1 asks for a position--time graph in which position
increases throughout the plotted interval. The slope need not
be constant. The original rubric specifies one nominal condition,
with a maximum score of 4 points:
\begin{itemize}
    \item Increasing position: \(dx/dt > 0.05\) throughout
    the plotted interval (4 points).
\end{itemize}
The straight line in Fig.~\ref{fig:ex1} is one possible
reference answer.

\begin{figure}
\centering
\includegraphics[width=\columnwidth]{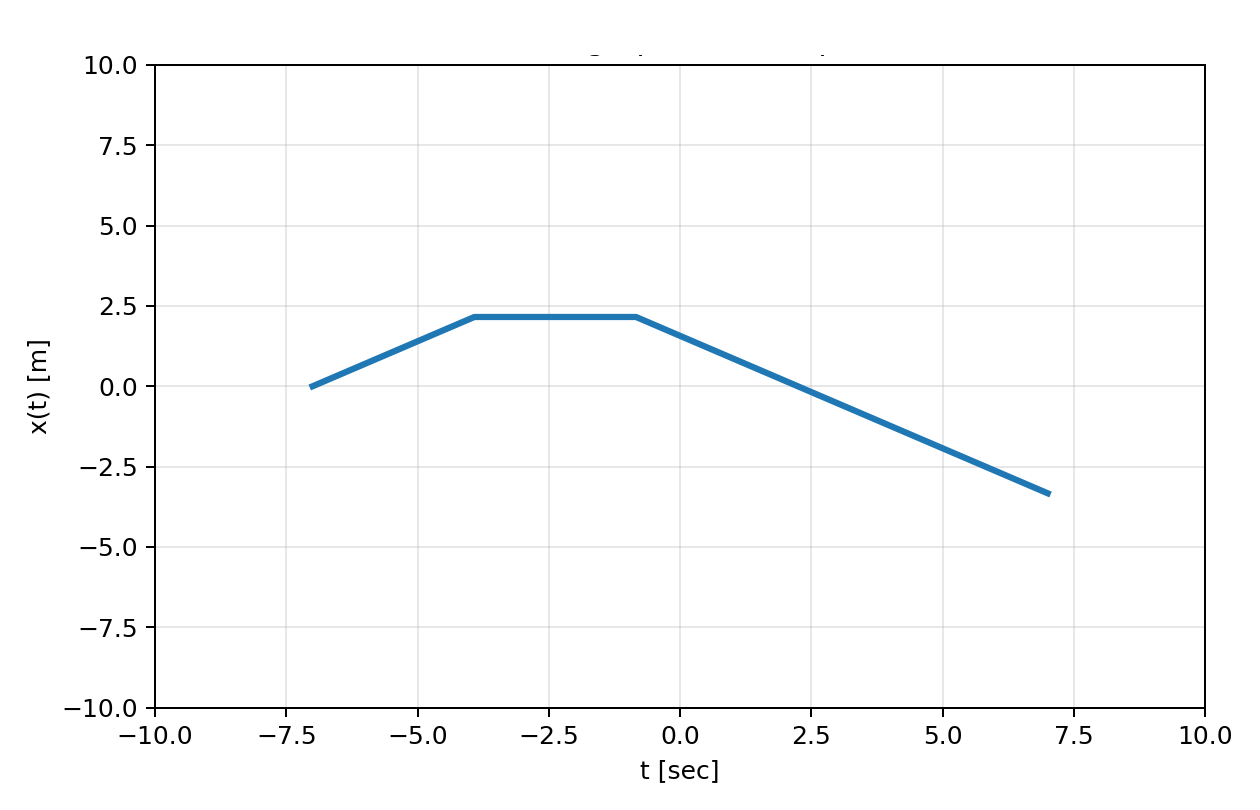}
\caption{One reference graph for Exercise 2, with rising, horizontal, and falling segments representing forward motion, rest, and backward motion.}
\label{fig:ex2}
\end{figure}

\paragraph*{Ex.\,2, Forward / stop / reverse.}
Exercise 2 asks for a position--time graph of a car that moves
forward, stops for a few seconds, and then moves backward.
The original rubric specifies three nominal conditions,
with a maximum score of 10 points; the instruction specifies
their order and the portions of the curve to which they apply:
\begin{itemize}
    \item Forward motion: \(dx/dt > 0.1\) during the initial
    forward phase (3 points).

    \item Stopping interval: \(dx/dt = 0\) over a finite
    interval between the moving phases (4 points).

    \item Reverse motion: \(dx/dt < -0.1\) during the subsequent
    backward phase (3 points).
\end{itemize}
An isolated point of zero slope does not satisfy the requirement
for a stopping interval. The piecewise-linear graph in
Fig.~\ref{fig:ex2} is one possible reference answer; constant
slopes during the moving phases are not required by the scenario.

\begin{figure}
\centering
\includegraphics[width=\columnwidth]{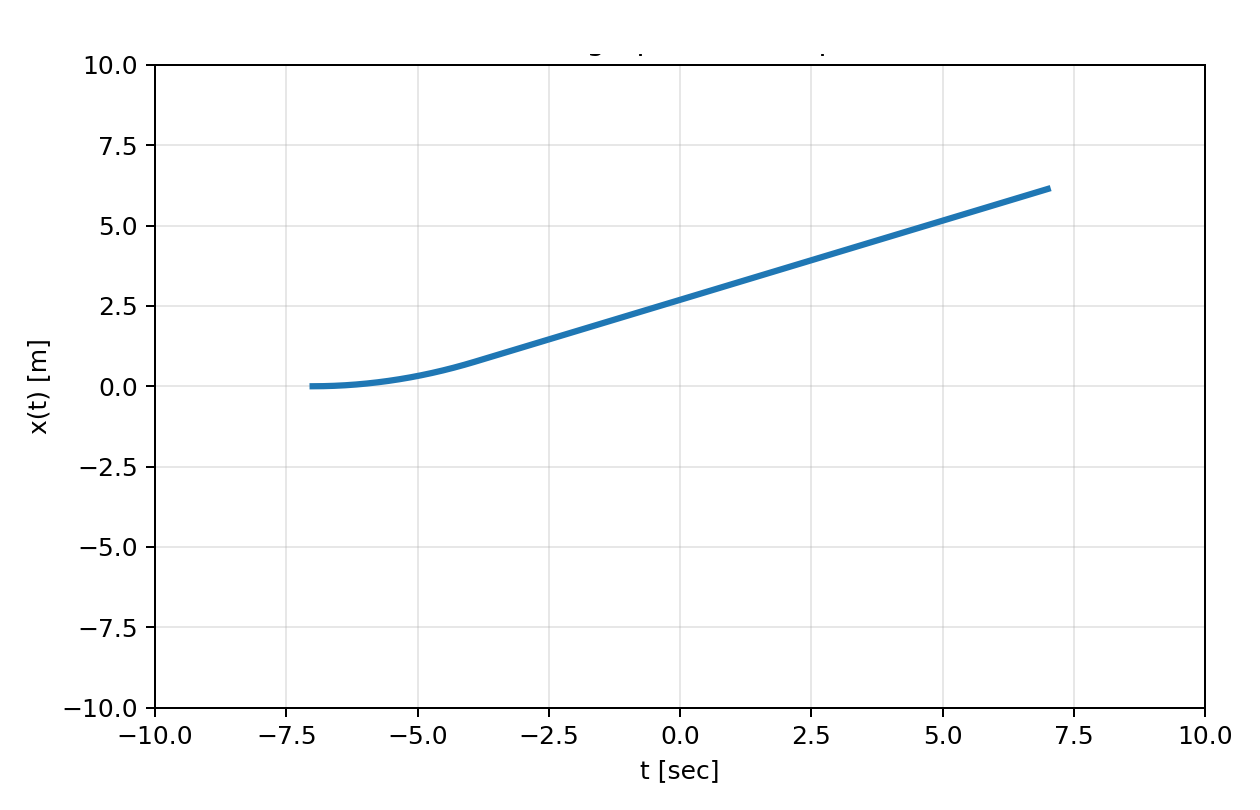}
\caption{One reference graph for Exercise 3. The curve starts with zero slope, bends upward during acceleration, and becomes linear during the subsequent phase of constant velocity.}
\label{fig:ex3}
\end{figure}

\paragraph*{Ex.\,3, Acceleration from rest followed by constant velocity.}
Exercise 3 asks for a position--time graph of a car that
accelerates forward from rest for a few seconds and then moves
with constant velocity. The acceleration need not be constant.
The original rubric specifies four nominal conditions,
with a maximum score of 10 points; the portion of the curve
to which each applies is given by the instruction:
\begin{itemize}
    \item Initial rest: \(dx/dt = 0\) at the beginning
    of the curve (2 points).

    \item Forward motion: \(dx/dt \ge 0\) throughout
    the plotted interval (2 points).

    \item Acceleration: \(d^2x/dt^2 > 0.05\) during the initial
    accelerating phase (3 points).

    \item Constant velocity: \(d^2x/dt^2 = 0\) during
    the subsequent phase (3 points).
\end{itemize}
Figure~\ref{fig:ex3} shows one possible reference answer.

\paragraph*{Ex.\,4, Forward / stop / reverse: velocity--time graph.}
Exercise 4 uses the same verbal motion scenario as Exercise 2
but asks for a velocity--time graph. The original rubric
specifies three nominal conditions, with a maximum score
of 10 points; the instruction specifies their order and
the portions of the curve to which they apply:
\begin{itemize}
    \item Forward motion: \(v > 0.1\) during the initial
    forward phase (3 points).

    \item Stopping interval: \(v = 0\) over a finite
    interval between the moving phases (4 points).

    \item Reverse motion: \(v < -0.1\) during the subsequent
    backward phase (3 points).
\end{itemize}
Constant velocity during the moving phases is not required.
In this representation, the sign of the plotted velocity
determines the direction of motion, while a sustained zero
value represents rest. Figure~\ref{fig:ex4} shows one possible
reference answer.

\subsubsection{Prompt for AI feedback}
\label{sec:prompt}

One prompt was used for each generation direction. Both requested
JSON output containing, for each rubric criterion, a satisfaction
flag, awarded points, and an explanation, together with the total
score. The model and scoring policy differed between directions.

\paragraph{Forward direction (Listing~\ref{lst:forward}).}
Forward submissions were graded by GPT-5.1 through Ethel using
Azure OpenAI. The prompt required full points or zero for each
criterion. It supplied the submitted image, the exercise
instruction, the rubric, and an exercise-specific interpretation
hint. These hints directed attention to slope in Exercises~1
and~2, slope and curvature in Exercise~3, and the plotted
velocity value in Exercise~4. Listing~\ref{lst:forward}
reproduces the original prompt used in the earlier
study~\cite{10.1007/978-3-032-41086-3_13}.

\paragraph{Reverse direction (Listing~\ref{lst:reverse}).}
Reverse submissions were graded by GPT-6 astra through the Azure
OpenAI deployment \texttt{Ethel\_6astra} (13 September 2026).
Each image was graded in a separate request, without earlier
images or responses. No temperature or token limit was specified
in the requests.

The standalone prompt supplied the submitted image, the exercise
instruction, and the rubric, without a reference graph or an
additional exercise-specific interpretation hint. The prompt
also included a step filename; a descriptive filename could
convey information about the generation stage.

The prompt requested partial credit scaled by closeness to each
criterion's numerical threshold, and near-zero credit for a
value with the wrong sign or far from the threshold. Fractional
awards are therefore permitted by this policy.

In both directions, the total awarded score is the sum of the
criterion-level awards, as defined in Eq.~\ref{eq:grading}.
\subsubsection{Grouping forward submissions}
\label{sec:classification}

Let $\mathcal{D}_i$ denote the set of submissions generated in the
forward direction for assignment $q_i$. For each submission $x_n$,
$D_{\mathrm{ref}}(x_n,a_i)$ denotes the cosine distance between
embeddings of $x_n$ and the reference graph $a_i$ obtained using
ResNet-50~\cite{Surrogate_He2016}. The normalized awarded score is
$\rho_G(x_n)$ (Eq.~\ref{eq:ratio}). Using a visual-distance cutoff
$\delta = 0.25$ and a score cutoff $\tau = 0.60$, we consider two
groups:
\begin{align}
\mathcal{D}_i^{\mathrm{conf}} &= \left\{ x_n \in \mathcal{D}_i :
D_{\mathrm{ref}}(x_n, a_i) < \delta,\ \rho_G(x_n) < \tau \right\},
\label{eq:conf}\\
\mathcal{D}_i^{\mathrm{rob}} &= \left\{ x_n \in \mathcal{D}_i :
D_{\mathrm{ref}}(x_n, a_i) \ge \delta,\ \rho_G(x_n) \ge \tau \right\}.
\label{eq:rob}
\end{align}

The groups are referred to as \emph{Confusing Images}
($\mathcal{D}_i^{\mathrm{conf}}$) and \emph{Robust-to-Noise Images}
($\mathcal{D}_i^{\mathrm{rob}}$), respectively. These are labels for
combinations of visual distance and awarded score; membership
does not by itself demonstrate confusion, robustness, or a
grading error.

Here, $\tau$ defines the score groups used in this analysis.
Neither visual distance from the reference nor crossing this
score cutoff independently establishes whether a submission
satisfies the physical criteria. Assessing the recorded award
requires comparing the submitted graph and accompanying
information with the exercise instructions and rubric.

An exploratory analysis of a differentiable score surrogate,
including its limited agreement with the grader and the resulting
attribution maps, is reported in Appendix~\ref{app:surrogate}.

\subsubsection{Visual summaries of awarded credit}
\label{sec:deterministic}

For each graded image $x^{(j)}$ in a reverse trajectory
(Eq.~\ref{eq:trajectory}), the recorded output includes the rule
identifiers, satisfied/not-satisfied flags, awarded points, and
explanations. Here, $j$ indexes the trajectory image and $k$
indexes a rubric criterion. Let $p_{kj}$ denote the points awarded
to criterion $k$ for image $x^{(j)}$. The corresponding awarded
fraction is
\begin{equation}
\hat{\varphi}_k\!\left(x^{(j)}\right)
= \frac{p_{kj}}{w_k}
\in [0,1].
\end{equation}
The total awarded score is the weighted sum in Eq.~\ref{eq:grading}.

A deterministic plotting routine displays the motion segments
associated with the rubric, annotated with these awarded fractions.
We refer to this display as a \emph{grade schematic}. It summarizes
the recorded awards; it does not uniquely reconstruct either the
submitted graph or the model's interpretation of that graph.

For selected examples, we examine the original image, the rubric,
the information supplied with the submission, and the recorded
explanation together to assess whether an award is physically
justified. Trajectories I and II and the samples in Appendix~\ref{sec:extra-samples}
were drawn at random (fixed seed) from the graded reverse-direction images;
the acceleration examples were selected to span the range of awarded credit. These examples are presented in
Secs.~\ref{sec:traj1}--\ref{sec:concavity}.

\section{Results}
\label{sec:results}

\subsection{Forward direction}
\label{sec:results-forward}
The forward analysis (Sec.~\ref{sec:forward}) grouped submissions
using visual distance from the reference and normalized awarded
score (Sec.~\ref{sec:classification}). Exercise~3, whose rubric
involves the second derivative $d^2x/dt^2$, had the largest
proportion of its graded submissions in the \emph{Confusing Images}
group. These submissions had visual distances below $\delta$ and
awarded score ratios below $\tau$. This pattern identifies cases
for inspection, but does not establish that curvature was assessed
incorrectly or inconsistently.

\begin{figure}
\centering
\includegraphics[width=\linewidth]{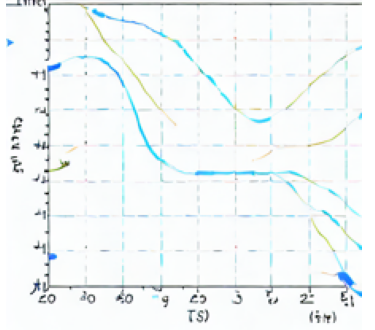}
\caption{A highly distorted version of the graph in Fig.~\ref{fig:ex4}.}
\label{fig:RTN4}
\end{figure}

Figure~\ref{fig:RTN4} shows an Exercise~4 submission in the
\emph{Robust-to-Noise Images} group that received the full
$10/10$ points. The image contains disconnected fragments
and apparent branchings, and its distorted axes and labels
do not establish a clear zero-velocity reference.
Table~\ref{tab:ex4} summarizes the criterion-level awards and
reported explanations. The grader awards all four stopping
points and states that the flat segment lies at $v=0$.
However, a horizontal velocity--time segment indicates
constant velocity; identifying rest additionally requires
that this value be zero. The displayed image therefore
does not unambiguously support the assertion used to award
full credit for the stopping criterion.

\begin{table*}
\caption{\label{tab:ex4}
Grader judgment for the highly distorted graph in Fig.~\ref{fig:RTN4}.
}%
\begin{ruledtabular}
\begin{tabular}{lll}
Criterion & Award & Grader's explanation (abridged) \\
\hline
Forward motion	&3/3&The velocity curve begins above zero.\\
Stopping interval	&4/4&	There is a sustained flat segment on the v = 0 line, indicating a stop.\\
Reverse motion	&3/3&	After the stop, the velocity values are below zero.\\
Total	&10/10&	All three criteria were marked satisfied.
\end{tabular}
\end{ruledtabular}
\end{table*}

Eventually, distortion can lead to loss of points, although
the present results do not establish a monotonic relation
between distortion and awarded score. Across the 300
forward-generated Exercise~4 submissions, the mean awarded
score was $8.16/10$, and at least 16 submissions received less
than $6/10$. The full-credit award in Fig.~\ref{fig:RTN4}
therefore occurred within a corpus in which substantial
deductions were also recorded. Appendix~\ref{sec:extra-samples}
shows additional examples from the reverse experiment,
including submissions awarded $5/10$ and $3/10$, under its
different model and scoring policy.

\subsection{Reverse generation}
\label{sec:results-backward}

\subsubsection{Examples}
\label{sec:traj1}

\begin{figure*}
\centering
\includegraphics[width=\linewidth]{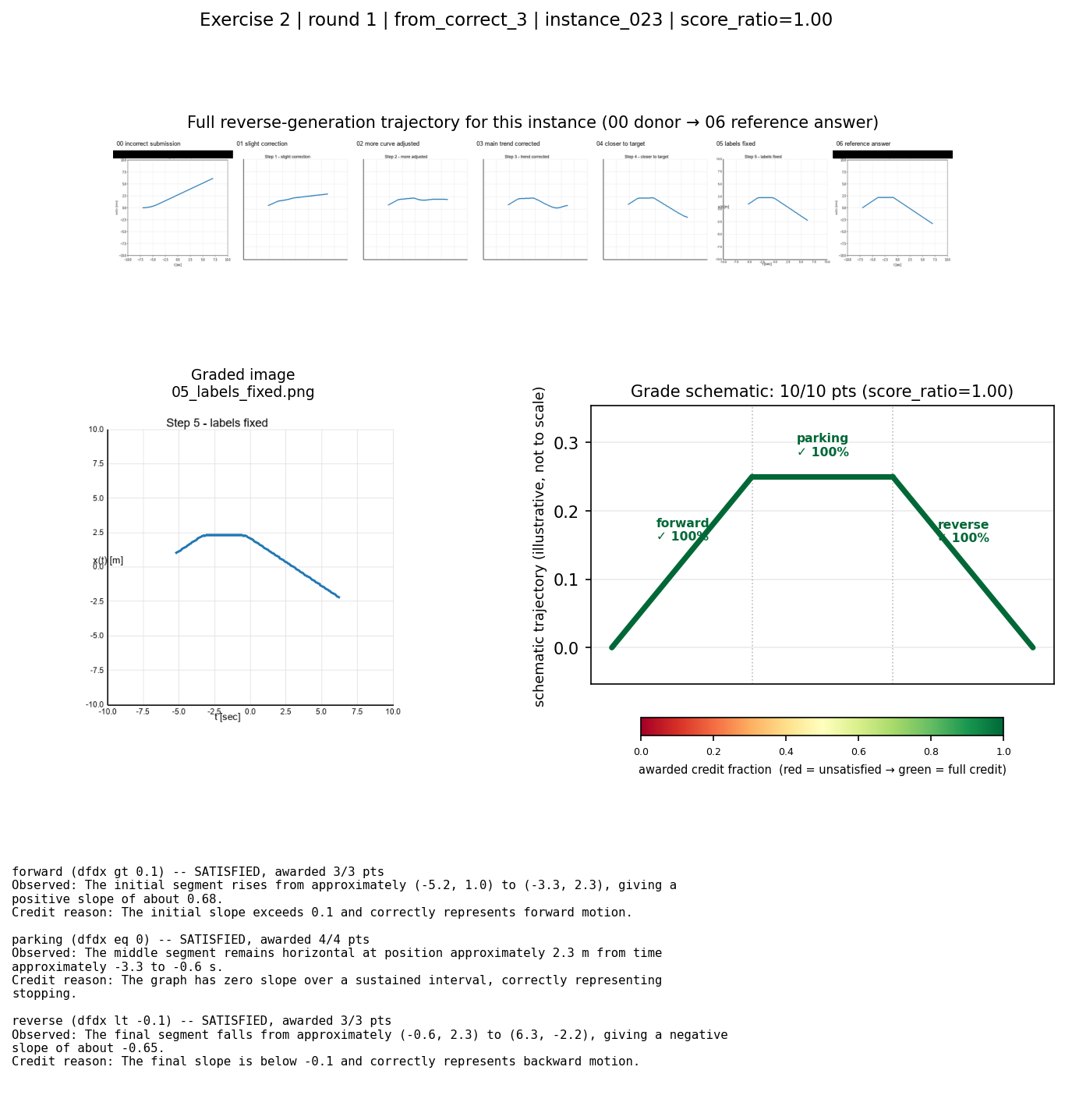}
\caption{Exercise 2, round 1, instance\_023 (source: Exercise 3), step 5, score\_ratio $=1.00$. Left: synthetic generated graph graded by the model. Right: grade schematic,
forward 100\%, parking 100\%, reverse 100\%.}
\label{fig:traj1}
\end{figure*}

\paragraph{Trajectory I.}
Figure~\ref{fig:traj1} presents a selected Exercise~2 submission
from a reverse-generation trajectory (Eq.~\ref{eq:trajectory}).
The left panel shows the submitted position--time graph, and the
right panel shows the corresponding grade schematic
(Sec.~\ref{sec:deterministic}). The schematic shows full credit
for all three rubric criteria: forward motion, a stopping
interval, and backward motion.

In the submitted position--time graph in Fig.~\ref{fig:traj1},
position first increases, remains approximately constant over a
finite interval, and then decreases. The recorded grading
explanation estimates an initial slope of approximately
$0.68\,\mathrm{m/s}$ and a final slope of approximately
$-0.65\,\mathrm{m/s}$, comparing these values with the encoded
cutoffs $0.1$ and $-0.1$, respectively. It identifies a horizontal
interval from approximately $t=-3.3\,\mathrm{s}$ to
$t=-0.6\,\mathrm{s}$, corresponding to about $2.7\,\mathrm{s}$
at rest. These estimates are consistent with the curve and
labeled axes in the submitted image. The full-credit award is
consistent with the required forward--stop--reverse motion,
and the grader's numerical comparisons can be checked using
the supplied axis scales.

\paragraph{Trajectory II.}
\label{sec:traj2}

\begin{figure*}
\centering
\includegraphics[width=\linewidth]{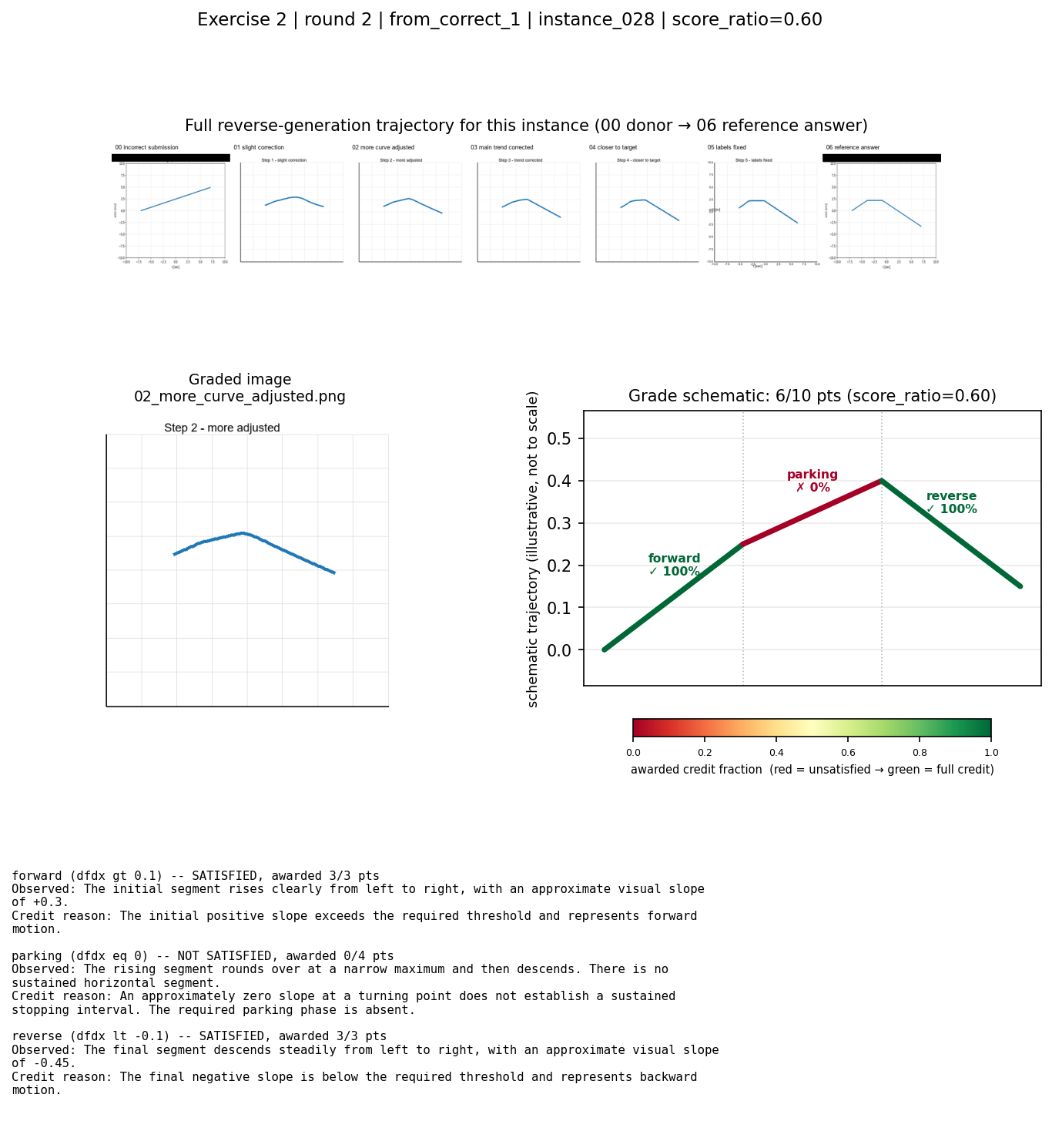}
\caption{Exercise 2, round 2, instance\_028 (source: Exercise 1), step 2, score\_ratio $=0.60$. Graded synthetic image and its grade schematic.}
\label{fig:traj2}
\end{figure*}

Figure~\ref{fig:traj2} presents a second Exercise~2 submission from a reverse-generation trajectory.

For the submission in Fig.~\ref{fig:traj2}, the recorded awards
are $3/3$ points for forward motion, $0/4$ for the stopping
interval, and $3/3$ for backward motion, for a total of $6/10$
(score ratio $0.60$). Because criteria are scored additively, the
submission reaches $\tau = 0.60$ although the stopping interval is absent.
The submitted position--time graph rises to a narrow maximum and
then falls, without a sustained horizontal segment.

The grader's explanation explicitly distinguishes a turning
point from a stopping interval. A sustained stop requires
constant position over a finite interval of time. Even if the
velocity vanishes at a single smooth turning point, this does
not establish such an interval. Withholding the four points for
the stopping criterion is therefore consistent with the
displayed graph.

The rising and falling portions are consistent with forward and
backward motion, respectively. The grader explicitly describes
its estimates of $+0.3$ and $-0.45$ as approximate visual slopes
and compares them with the encoded cutoffs of $0.1$ and $-0.1$.
Such a comparison could implement an operational criterion for
a visibly increasing or decreasing graph. However, the grading
instructions did not define the plotting coordinates in which
these cutoffs were to be applied. The explanation therefore
supports the qualitative interpretation of the motion without
establishing that the visual estimates use the intended
numerical convention. The distinction between a turning point
and a sustained stopping interval does not depend on that
convention.

\subsubsection{Assessment of the acceleration criterion}
\label{sec:concavity}

\begin{figure*}
\centering
\includegraphics[width=\linewidth]{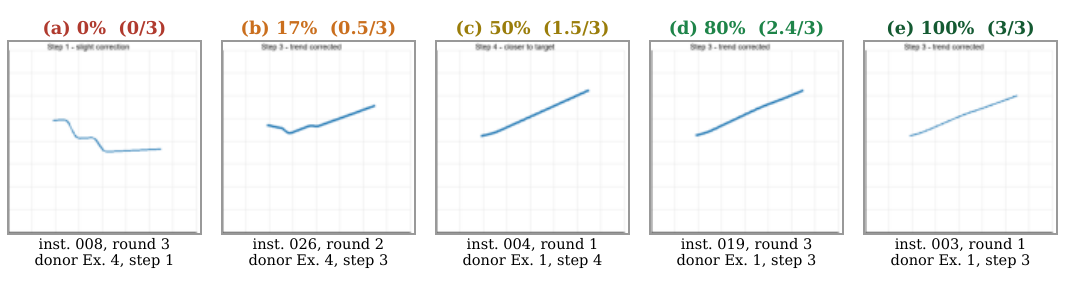}
\caption{Five selected Exercise~3 submissions, with the recorded
percentages of credit awarded for the acceleration criterion;
the accompanying grading explanations are given in
Table~\ref{tab:acc-explanations}. The criterion carries 3 points
and uses the encoded condition $d^2x/dt^2 > 0.05$. The numerical
cutoff was supplied without attached units or a defined plotting
normalization.}
\label{fig:acc-montage}
\end{figure*}

Figure~\ref{fig:acc-montage} presents five selected Exercise~3
submissions from different reverse-generation trajectories,
together with their recorded awards for the acceleration
criterion; the accompanying grading explanations are given in
Table~\ref{tab:acc-explanations}. The
displayed percentages indicate the fraction of the 3 available
points awarded for this criterion.

The implemented acceleration criterion uses the numerical
condition $d^2x/dt^2 > 0.05$ during the initial accelerating
phase. Positive curvature in a position--time graph supports
the qualitative requirement of positive acceleration. Applying
the encoded cutoff additionally requires a coordinate scale or
an explicit plotting convention. The grading instructions did
not specify such a convention for unnumbered graphs, so the
qualitative judgment and the numerical comparison must be
considered separately.

\begin{table*}
\caption{Saved grading explanations for the acceleration criterion of the
five submissions in Fig.~\ref{fig:acc-montage}, reproduced verbatim from the
recorded outputs (reverse direction, Listing~\ref{lst:reverse}).}
\label{tab:acc-explanations}
\begin{ruledtabular}
\small
\begin{tabular}{@{}llll@{}}
Panel & Credit & Observed & Reason for credit \\
\hline
(a) & 0\% (0/3) & \parbox[t]{0.42\textwidth}{\raggedright After the initial plateau, the slope becomes increasingly negative, producing concave-down motion. Later concave-up recovery portions flatten the downward motion rather than establish the requested initial forward acceleration.} & \parbox[t]{0.34\textwidth}{\raggedright The initial acceleration phase has negative curvature, the wrong sign for the required positive second derivative.} \\
\addlinespace
(b) & 17\% (0.5/3) & \parbox[t]{0.42\textwidth}{\raggedright The early descending segment initially steepens downward. Concave-up bending appears locally around the trough, but the subsequent rise flattens briefly before joining the final line. Unnumbered axes prevent a calibrated numerical curvature measurement.} & \parbox[t]{0.34\textwidth}{\raggedright Positive curvature is limited to local portions rather than sustained throughout the first phase. Much of the early curvature has the wrong sign or is approximately zero.} \\
\addlinespace
(c) & 50\% (1.5/3) & \parbox[t]{0.42\textwidth}{\raggedright There is a modest increase in slope near the beginning, from a shallow rising segment to the steeper main segment. Positive curvature is concentrated around this short bend; much of the initial portion looks nearly straight. Unnumbered axes prevent a reliable numerical curvature measurement.} & \parbox[t]{0.34\textwidth}{\raggedright The early slope change provides evidence of positive acceleration, but the visible concavity is limited and does not clearly establish the required curvature throughout the first phase.} \\
\addlinespace
(d) & 80\% (2.4/3) & \parbox[t]{0.42\textwidth}{\raggedright The beginning bends upward as its slope increases from shallow to moderately positive. The curvature is modest and becomes difficult to distinguish from a straight line. Unnumbered axes prevent a calibrated numerical second-derivative measurement.} & \parbox[t]{0.34\textwidth}{\raggedright Positive initial curvature is visible, earning substantial credit, but the evidence for exceeding the specified curvature threshold is limited.} \\
\addlinespace
(e) & 100\% (3/3) & \parbox[t]{0.42\textwidth}{\raggedright The early section bends upward: its slope increases from roughly 0.25 to roughly 0.45 before the later straight portion. The curvature is consistent with exceeding 0.05 when estimated using the displayed grid spacing, although numerical axis scales are absent.} & \parbox[t]{0.34\textwidth}{\raggedright The first phase visibly has increasing positive slope, providing the required positive second derivative.} \\
\end{tabular}
\end{ruledtabular}
\end{table*}

In Fig.~\ref{fig:acc-montage}, the submission receiving 0\% of
the acceleration credit begins with a nearly horizontal segment
that turns downward. During this transition, the slope
decreases from approximately zero to negative values,
corresponding to negative acceleration. The grader's displayed
explanation identifies this sign mismatch, which supports
withholding credit for the required initial positive acceleration.

The submission receiving 17\% contains an initial dip, a local
concave-up region near the minimum, and a brief flattening of
the subsequent rise. Its explanation emphasizes that positive
curvature is confined to portions of the curve. The submissions
receiving 50\%, 80\%, and 100\% instead show rising curves with
an initially increasing slope followed by more nearly linear
portions. These examples therefore differ in physically
relevant features.

For the submission receiving 100\%, the explanation reproduced
in Table~\ref{tab:acc-explanations} reports an increase in slope
from roughly 0.25 to 0.45 and identifies positive curvature.
This qualitative observation is consistent with positive
acceleration. The explanation also invokes the displayed grid
spacing when comparing the curvature with the cutoff $0.05$.
The grading instructions, however, did not specify how grid
spacing should define the numerical scale for this comparison.
The explanation therefore applies a plotting convention that
was not established in the prompt.

Across these submissions, the sign and extent of the visible
curvature provide a basis for qualitative assessment.
Interpreting the numerical comparisons additionally requires
attention to the coordinate convention used to apply the
encoded cutoff. The different recorded awards are for
different inputs and do not establish score instability
across repeated gradings of the same response.

\begin{figure*}
\centering
\includegraphics[width=\linewidth]{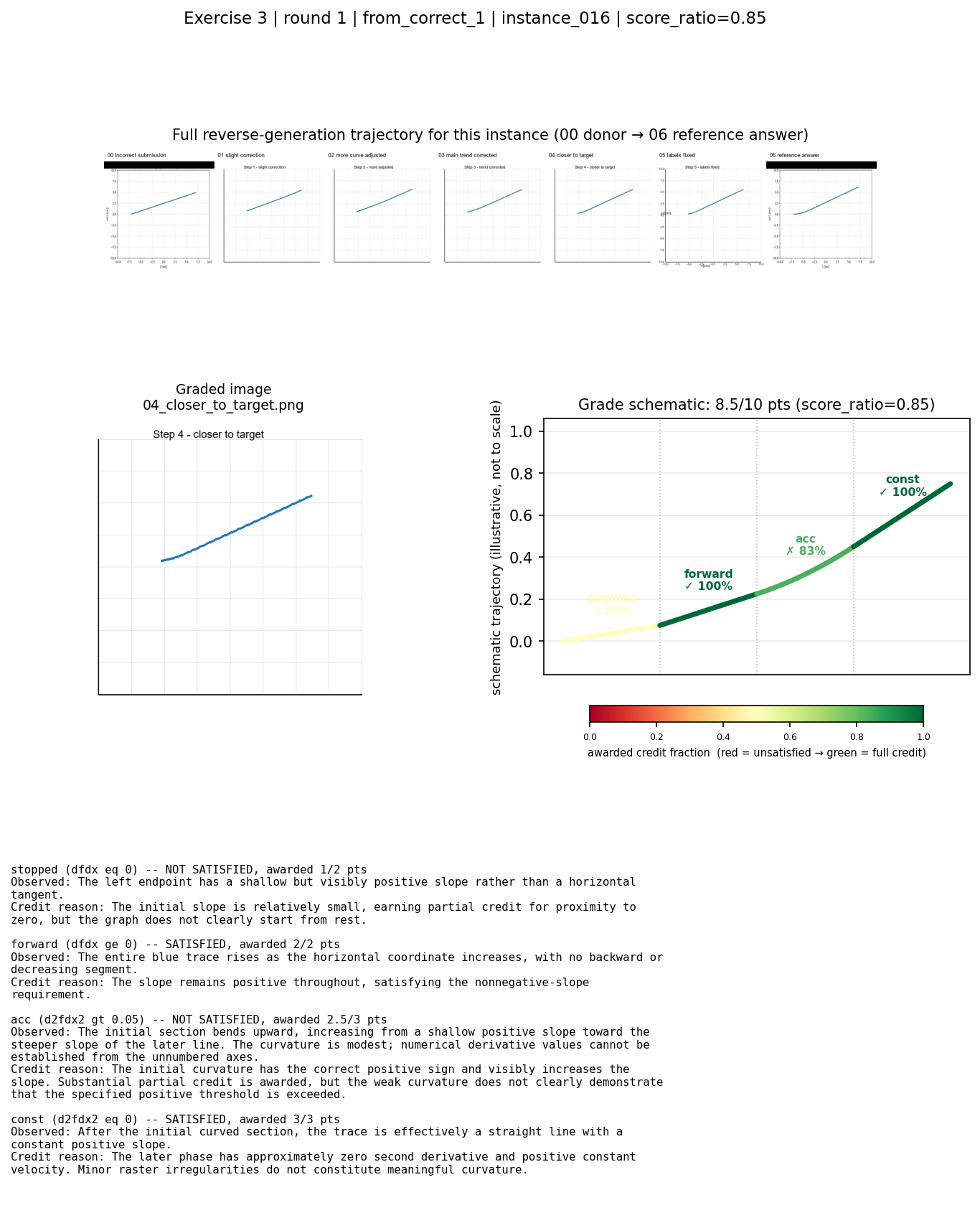}
\caption{Exercise 3, round 1, instance\_016 (source: Exercise 1), step 4, score\_ratio $=0.85$.
Acceleration: 2.5/3 points, marked not satisfied; initial rest: 1/2 points, marked not satisfied.}
\label{fig:inst016}
\end{figure*}

Figure~\ref{fig:inst016} shows an additional Exercise~3 submission
graded with the standalone reverse-direction prompt
(Listing~\ref{lst:reverse}). The graph begins with an upward bend
and continues as a nearly straight line. The grader identifies
positive curvature, but states that numerical derivative values
cannot be established from the unnumbered axes. It awards
$2.5/3$ points for acceleration and $1/2$ for initial rest,
marking both criteria as not satisfied. For the later
constant-velocity phase, it awards $3/3$ and states that minor
raster irregularities do not constitute meaningful curvature.
The total award is $8.5/10$.

The standalone policy permits partial credit according to
closeness to each criterion's numerical threshold. The output
includes visual judgments compatible with distinguishing
intended graph features from rendering irregularities. However,
it does not specify the numerical convention connecting the
observed curvature to the cutoff $0.05$, or how numerical
closeness determines the award of $2.5/3$ points. This limits
interpretation of the fractional award without establishing
that it is incorrect. The submitted image, the prompt, and
the saved output for this case are retained with the code.

\section{Discussion}
\label{sec:findings}
The main issue raised by these examples is how to distinguish
useful tolerance of imperfect drawings from an inadequately
supported award of credit. Several selected outputs identify
physical features correctly and assign points consistently
with those features. A high score on a degraded image,
however, can have different interpretations: the relevant
physics may remain clearly represented, or the grading
explanation may supply information that the image does not
establish. Robustness in physics assessment therefore requires
attention to which physical relationships are preserved and
which have changed or become ambiguous. This distinction matters for teaching because graphs are used
to connect mathematical representations with physical situations
and to develop physical reasoning~\cite{Redish_2023_Graphs}.

Figure~\ref{fig:RTN4} illustrates why apparent tolerance of
distortion can be troubling. Despite the severely degraded
image, the grader describes the required motion and awards
full credit. The image contains disconnected fragments and
apparent branchings, with competing traces that do not
unambiguously specify a single velocity at each time for
the car. These structural ambiguities, together with the
corrupted axes and labels, produced no deduction in the
reported award. The explanations summarized in
Table~\ref{tab:ex4} do not address these ambiguities.
In an informal follow-up discussion of
Fig.~\ref{fig:RTN4} with ChatGPT (GPT-6 Astra), the initial
critique emphasized the ambiguous zero-velocity reference
without addressing the disconnected and branching traces.
Those structural ambiguities were addressed only after a
follow-up prompt explicitly drew attention to them. Although this exchange was not a controlled evaluation,
it illustrates that an AI asked to review a grading decision
may also fail to point out the same visible problems in
the submitted graph.

The assertion that the flat segment lies at $v=0$ introduces
a further problem: horizontality establishes constant velocity,
but identifying rest additionally requires a zero-velocity
reference. The displayed image does not clearly establish that
reference. Selecting a dominant trace and assigning zero to
its intermediate plateau could reproduce the grader's
description, but the reported explanation does not justify
those choices. The full-credit award therefore rests on a
particular interpretation of the damaged image that is not
clearly supported by the submission.

Figure~\ref{fig:RTN4} also illustrates why automated grading
can conflict with an instructor's intuitive expectations.
An instructor might first ask whether the fragmented image
can be interpreted as a coherent graph of the requested
motion at all, before considering individual rubric points.
The recorded response nevertheless describes the required
features and awards full credit without addressing this
prior problem. Instructors should therefore not assume
that a submission they regard as too ambiguous for meaningful
assessment will elicit rejection or uncertainty from the
model. Stress testing should include such cases, since
apparently obvious defects in a physical representation
may have little effect on the awarded score.

A related issue arises in human grading: instructors sometimes
attribute correct reasoning to solutions that do not explicitly
show it~\cite{Henderson_2004}. The present cases similarly require
distinguishing evidence in the submission from an interpretation
supplied by the grader.

The recorded scores also show that the tested configurations
did not award full credit to every submission. Reduced scores
occurred within the forward corpus, while the reverse
examples include deductions with a clear physical basis.
In Fig.~\ref{fig:traj2}, the grader correctly distinguishes
zero slope at a turning point from rest over a finite
interval. This is a meaningful distinction for assessing
kinematics. It also illustrates why a total score provides
an incomplete account of a grading decision: the submission
reaches the chosen acceptance cutoff despite receiving no
credit for the required stopping interval. Reaching that
cutoff reflects the chosen scoring policy; it does not
certify that every physical requirement has been met. Examining individual awards also serves a diagnostic purpose:
separate rubric scores can identify specific difficulties
that an overall grade does not reveal, as demonstrated in
assessment of physics problem solving~\cite{Docktor_2016}.

The numerical cutoffs raise a related question about the
translation of an instructor's intentions into grading
instructions. In these qualitative exercises, the cutoffs
were intended to distinguish clearly expressed features
from minor drawing irregularities. Supplying the numbers
without this purpose leaves open interpretations such as
a threshold in plotting coordinates or a minimum physical
magnitude. The visual-slope estimates in
Fig.~\ref{fig:traj2} and the use of grid spacing in the
full-credit explanation associated with
Fig.~\ref{fig:acc-montage} illustrate how numerical conventions
can enter the reported justification. Interpreting the
numerical rule thus becomes part of the grading task,
alongside interpreting the graph. The rubric should
communicate both the intended physical requirement and
the coordinate convention used to make its numerical
implementation meaningful. This attention to coordinate conventions connects the grading
task to a broader educational objective: interpreting how
coordinate choices affect motion graphs~\cite{Klein_2017}.

Partial credit makes this distinction consequential for the
final score. A nearly horizontal position--time segment could
reflect an imperfect drawing of rest or a deliberate indication
of motion. The appropriate judgment depends on the exercise
requirements and the stated tolerance for imperfect drawing.
Likewise, identifying positive curvature does not by itself
determine a particular fraction of the acceleration points.
In Fig.~\ref{fig:inst016}, the explanation supports recognition
of that feature but does not establish how its numerical
proximity to the cutoff determines the fractional award.
Uncertainty about reading an image and partial satisfaction
of a physical requirement are different bases for withholding
some credit. A grading policy should state how each is to
be treated.

The practical implication is that instructors should examine
whether criterion-level awards are supported by the physical
information in the submitted graph.
Section~\ref{sec:outlook} outlines how this review could be
supported by an integrated system.

\section{Limitations}
This study uses synthetic graphs for four exercises. The graphs and reverse generation sequences have not been validated as representative of real student answers or revisions. Moving closer to a reference graph does not necessarily mean becoming more physically correct. The selected examples cannot be used to estimate how often the observed behavior occurs in real assessments.

Visual similarity is not a measure of physical correctness, and a passing total does not imply that every rubric criterion was met. Tick values and axis labels are drawn only on step 5; steps 1--4 show gridlines only. Qualitative features such as increasing position, linearity,
a sustained horizontal position--time interval, and positive
curvature can be assessed without numerical axis values.
However, the grading instructions did not define a coordinate
convention for applying the numerical cutoffs to the unnumbered
graphs. This limits the interpretation of numerical comparisons
and the resulting partial-credit awards. Score differences can therefore reflect the information supplied, the rubric, or the grader. The presented analysis does not compare all scores with an independent human reference, so it does not provide an overall grading accuracy or error rate.

The grade schematics display the credit awarded by the grader.
They do not reveal the MLLM's internal reasoning or independently
confirm that its scores are correct. The findings are limited to
the models, prompts, rubrics, and synthetic inputs examined.
The examples also do not establish score stability across repeated
runs. These results cannot establish general grading reliability
or benefits for student learning.

\section{Outlook}
\label{sec:outlook}

\begin{figure}
\centering
\includegraphics[width=\linewidth]{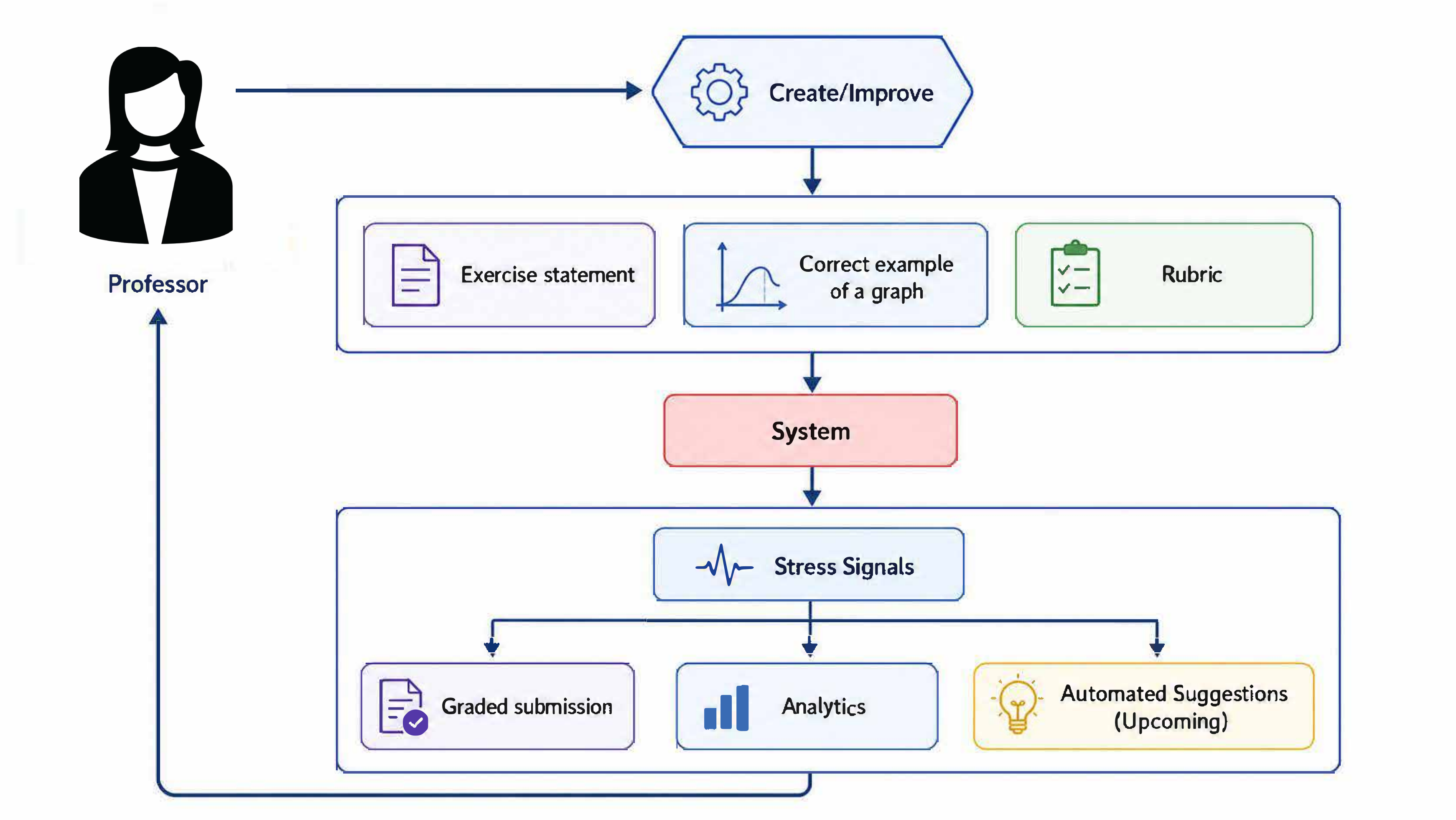}
\caption{Proposed workflow for instructor review of graphical
grading. Generated responses, criterion-level awards, and
diagnostic summaries support decisions about the exercise
and grading policy. The reference graph supports input
generation and is not supplied to the grader. Automated
suggestions are a prospective extension; the integrated
review and revision workflow has not been evaluated in
this study.}
\label{fig:overview}
\end{figure}

Figure~\ref{fig:overview} outlines a possible system for
reviewing graphical grading before classroom use. An
instructor would supply the exercise, rubric, and an
illustrative reference graph. The system would generate
varied responses, grade them with the intended grading
configuration, and present selected images alongside their
criterion-level awards and explanations. The reference
would guide generation without being supplied to the grader.

The present cases suggest concrete review questions:
does the grader distinguish a stopping interval from a
turning point, identify the zero-velocity reference on a
velocity--time graph, and apply numerical cutoffs and
partial-credit rules as intended? High-scoring responses
would need inspection alongside deductions, as
Fig.~\ref{fig:RTN4} illustrates. The instructor could then
clarify the exercise or grading policy and repeat the
grading to examine the consequences.

Reviewed examples and instructor judgments could be retained
for checks after changes to the rubric, prompt, or model.
Such checks would examine whether the revised awards reflect
the intended physics, including whether physically adequate
responses still receive credit. Automated suggestions could
propose revisions, with their adoption remaining an
instructor decision.

\section{Conclusion}

This exploratory study examined selected grading outputs for
synthetic responses to four kinematics exercises. Regarding
Research Question~\ref{rq:physical-interpretation}, the examples
included physically appropriate interpretations of slope and
curvature and a clear distinction between instantaneous zero
velocity and rest over a finite interval. Some criterion-level
awards were supported by these interpretations and by the
submitted graphs. The results therefore include examples of
grading grounded in the physical properties required by the
exercises.

Regarding Research Question~\ref{rq:graphical-evidence}, the
severely distorted velocity--time graph received full credit
despite disconnected and branching traces and an unclear
zero-velocity reference. The reported explanation did not
establish how these ambiguities were resolved. Substantially
lower scores elsewhere in the same forward corpus show that
this configuration did not invariably award full credit.
Nevertheless, the example demonstrates that a full-credit
award can rest on an interpretation insufficiently supported
by the displayed response. Evaluating tolerance of distortion
therefore requires checking whether the information needed
for the physical judgment remains identifiable.

Regarding Research Question~\ref{rq:numerical-criteria}, the
nonzero cutoffs were intended to distinguish clearly expressed
qualitative features from drawing irregularities. The grading
instructions did not communicate this purpose or define a
coordinate convention for applying the cutoffs to unnumbered
graphs. The reported explanations invoked visual slopes,
grid spacing, and qualitative curvature judgments, sometimes
in ways compatible with the intended role of the cutoffs.
However, they did not establish a common numerical convention
or how numerical closeness determined the exact partial-credit
awards. The reverse policy permitted fractional credit;
the unresolved numerical basis does not establish that those
fractions were incorrect.

The examined configurations demonstrate useful abilities to interpret kinematics graphs, but we would not recommend accepting their scores as final grades without instructor review. A grader can recognize the intended motion while overlooking defects that undermine the submitted graph as a physical representation. Apparent tolerance of distortion can therefore conceal unsupported credit. Instructors should not assume that a graph they regard
as uninterpretable will be rejected by the model. 

Clearer grading instructions may address some of these problems, but their effectiveness was not established here. The immediate value of stress testing is to expose such failures before students' grades depend on the system. For the configurations studied, scrutiny must include full-credit responses: the ability to recognize the intended physics does not yet justify trusting the grader to determine whether a submitted answer adequately represents it.

\acknowledgments
This work was supported by UID/00027/2025--Artificial Intelligence and Computer Science Laboratory (LIACC), with DOI \url{https://doi.org/10.54499/UID/00027/2025}, funded by Fundação para a Ciência e a Tecnologia, I.P./MECI through national funds, and by the Agenda ``Center for Responsible AI'', nr. C645008882-00000055, investment project nr. 62, financed by the Recovery and Resilience Plan (PRR) and by the European Union--NextGenerationEU.
The authors thank the ETH AI Center for its support and contribution to the interdisciplinary research context of this work. ChatGPT (GPT-6 Astra) assisted with manuscript revision, identification of relevant literature, and critical examination of the physical interpretations. The authors directed and
reviewed this assistance and take responsibility for the final text and scientific conclusions.

\section*{Data Availability}

The code for this study is available at \url{https://github.com/zammelRocks/kinematics-graph-grading-StressTesting} (reverse direction) \textit{[permanent location TBD]} and \url{https://github.com/zammelRocks/AnotherBrickInTheWall } (forward direction)
\textit{[permanent location TBD]}. The data for this study is available \textit{[permanent location TBD]}.
\appendix


\section{Image distortion prompt}
\label{app:img_prompt}

\subsection{Distorted image generation}
\begin{lstlisting}[caption={Image generation prompt used in the forward direction. The Stable Diffusion prompt is generated programmatically from the rubric, extracted by exercise rather than being one manually written prompt.
},
label={lst:img_forward},
basicstyle=\ttfamily\small,
breaklines=true,
frame=single,
columns=fullflexible]
"hand-drawn student sketch of a physics graph, pencil on graph paper, slight imperfections, natural handwriting jitter"

For Exercise 4, the rubric specifies:
- velocity > 0.1 for the forward segment
- velocity = 0 during parking
- velocity < -0.1 for the reverse segment

The generated Stable Diffusion prompt is:

"hand-drawn student sketch of a physics graph, pencil on graph paper, slight imperfections, natural handwriting jitter; the curve's value clearly above 0.1; the curve's value flat at 0; the curve's value clearly below -0.1"
\end{lstlisting}

\section{Grading prompts}
\label{app:prompt}

\subsection{Forward direction}
\begin{lstlisting}[caption={Grading prompt used in the forward direction.
The fields are filled from each exercise's configuration.},
label={lst:forward},
basicstyle=\ttfamily\small,
breaklines=true,
frame=single,
columns=fullflexible]
You MUST output ONLY valid JSON.

Instruction:
{instruction}

Rubric:
{rubric}

Interpretation guidance:
{interpretation_hint}

GENERAL RULES:
- Evaluate EACH rule independently
- Assign full points or zero per rule (no partial credit within a rule)
- Provide a concise, scientific explanation per rule

Output format:
{
  "total_score": number,
  "max_score": {max_score},
  "rules": [
    { "rule_id": "...", "satisfied": true/false,
      "points_awarded": number, "max_points": number,
      "explanation": "..." }
  ]
}
\end{lstlisting}

\subsection{Reverse direction (standalone policy)}
\begin{lstlisting}[caption={Grading prompt used in the reverse direction.
The system message and user message are shown in order; the submitted
image is attached to the user message.},
label={lst:reverse},
basicstyle=\ttfamily\small,
breaklines=true,
frame=single,
columns=fullflexible]
[System message]
You are a strict grading assistant for kinematics graph images. You grade
according to explicit rubric checkpoints and partial credit. You always
return valid JSON only.

[User message]
You are a strict grading assistant for kinematics graph images.

You will receive:
1. A submitted graph image, to grade exactly as a complete, standalone answer.
2. The exercise instruction.
3. The exact JSON rubric.

Your task:
Grade the submitted image as if it were a student's complete, final answer
to this exercise. You are not told, and must not assume, whether this image
was produced as part of some other process, a draft, or a revision -- judge
it purely on its own merits against the instruction and rubric below,
exactly as you would any standalone submission.

Partial-credit policy (applies uniformly to every rule):
- Each rule specifies a quantity, a relation, and an expected threshold
  (for example "dfdx gt 0.1").
- Award partial credit ONLY based on how close the image's observed value
  for that quantity is to satisfying its own expected threshold -- never
  based on assumed effort, assumed intent, or assumed progress toward some
  other image you have not been shown.
- A rule whose observed value is on the correct side of the threshold but
  has not yet reached it may receive partial credit, scaled by how close
  the observed value is to that threshold.
- A rule whose observed value has the wrong sign, or is far from its
  threshold, must receive near-zero credit.
- Penalize any extra non-graph artifacts, hooks, side fragments, unrelated
  lines, wrong direction, wrong function family, wrong slope sign, wrong
  axes, or missing labels if labels are expected.

Exercise number: {exercise_number}

Exercise instruction:
{instruction}

Rubric JSON:
{rubric_json}

Rule interpretation guidance:
- "points" is the maximum for that rule.
- "awarded" must be between 0 and that rule's "points".
- "total_awarded" must equal the sum of all rule awarded values.
- "score_ratio" must equal total_awarded / max_score.
- "score_100" must equal round(score_ratio * 100).
- Preserve all rubric rule ids.
- Preserve each rule's points, quantity, relation, and expected values
  in your output.
- For "observed", describe the visual evidence from the generated image.
- For "partial_credit_reason", explain why the rule got that many points.

Return ONLY valid JSON, no markdown, no explanations outside JSON.

Required JSON shape:
{
  "max_score": {max_score},
  "total_awarded": 0,
  "score_ratio": 0.0,
  "score_100": 0,
  "step_file": "filename.png",
  "rules": [
    { "id": "...", "points": 0, "awarded": 0, "quantity": "...",
      "relation": "...", "expected": 0,
      "observed": "describe what you see in the generated image",
      "satisfied": false,
      "partial_credit_reason": "explain awarded points" }
  ],
  "feedback": "short global feedback",
  "is_graph_only": true
}

--- This specific image ---
Step file: {step_file}

Grade the image described above now, following the general grading policy
stated earlier in this prompt.
\end{lstlisting}

\section{Exploratory surrogate attribution analysis}
\label{app:surrogate}

As an auxiliary analysis, we examined whether a differentiable
surrogate could approximate the grader's normalized scores and
support image-attribution visualizations. The surrogate was fitted
to image--score pairs from the forward-generated corpus, using the
recorded normalized score $\rho_G(x)$ defined in
Eq.~\ref{eq:ratio} as the prediction target. The model comprised a
ResNet-50 encoder~\cite{Surrogate_He2016} and a linear output head.
Fine-tuning of the encoder was restricted to its final residual
block (\texttt{layer4}), and fitting used a squared-error loss.

The reported evaluation used a held-out set of 60 images.
Table~\ref{tab:surrogate-agreement} summarizes the agreement between
surrogate predictions and the grader's recorded normalized scores.
The correlations indicate limited agreement. These measures assess
reproduction of the grader's scores; they do not measure agreement
with an independent assessment of the physics.

\begin{table}
\caption{Reported agreement between surrogate predictions and
recorded normalized grader scores on 60 held-out images.}
\label{tab:surrogate-agreement}
\begin{ruledtabular}
\begin{tabular}{lr}
Measure & Value \\
\hline
Mean squared error & 0.0372 \\
Root mean squared error & 0.193 \\
Pearson correlation & 0.28 \\
Spearman rank correlation & 0.36 \\
\end{tabular}
\end{ruledtabular}
\end{table}

\begin{figure*}[t]
\centering
\includegraphics[width=0.9\textwidth]{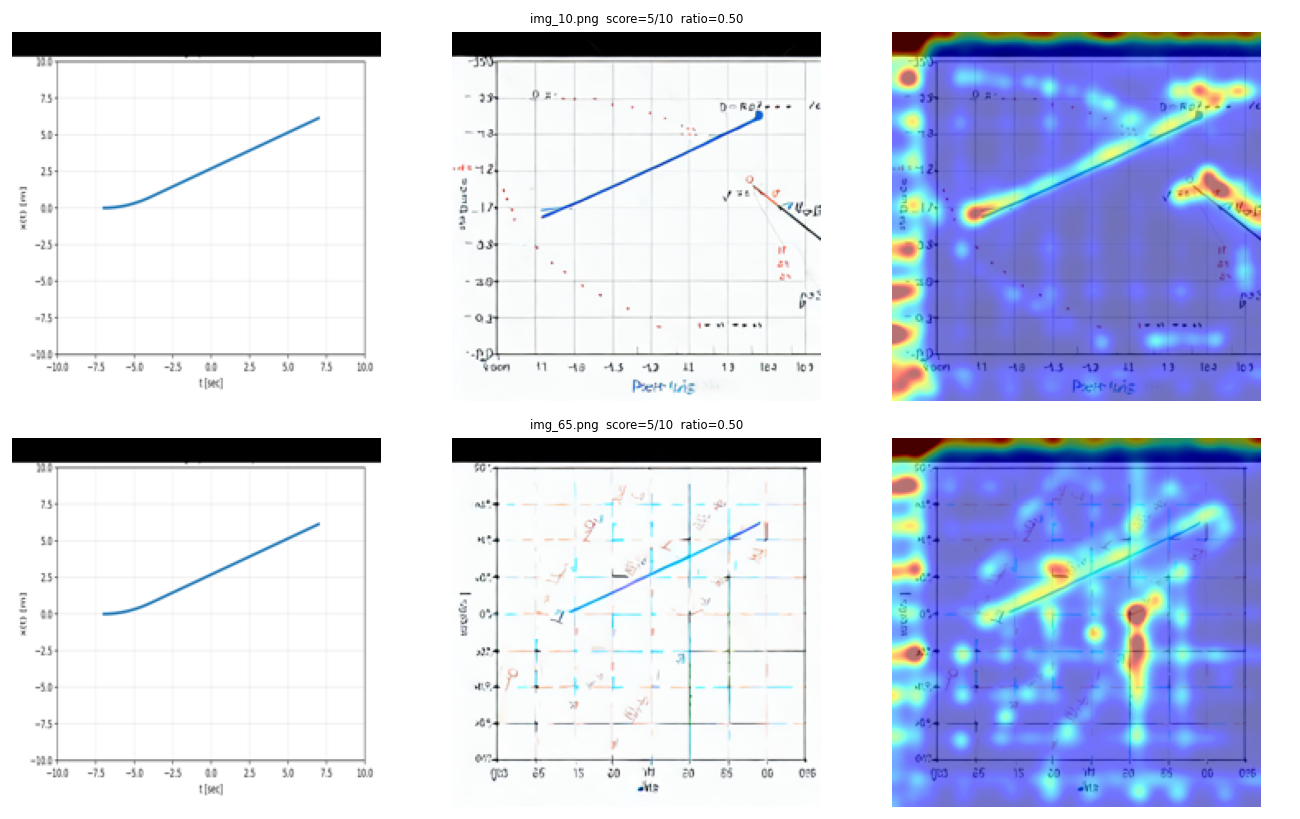}
\caption{Selected synthetic submissions for Exercise 3 and
attribution maps of the score surrogate. Each row shows, from
left to right, the reference graph, the submitted image, and
the map obtained using Integrated Gradients with SmoothGrad.
The maps describe the surrogate under this attribution procedure;
they do not establish which image features the MLLM grader used.}
\label{fig:surrogate-attributions}
\end{figure*}

Image attributions were calculated on the surrogate using
Integrated Gradients~\cite{Surrogate_Sundararajan2017} combined
with SmoothGrad~\cite{Surrogate_Smilkov2017}, and
Grad-CAM~\cite{Surrogate_Selvaraju2017}.
Figure~\ref{fig:surrogate-attributions} presents selected maps
obtained with Integrated Gradients and SmoothGrad for Exercise 3.
In these examples, attribution patterns appear along portions of
the plotted curve and around other image features, including the
plotting frame and added marks.

The limited score agreement constrains the interpretation of these
visualizations. Even close agreement in predicted scores would not
by itself establish that the surrogate and the MLLM depend on the
same image features. We therefore do not use these maps to infer
the MLLM grader's attention, internal reasoning, or the physical
correctness of its awards. The analysis of kinematics judgments
in the main text is based on the submitted graphs, exercise
criteria, and recorded grading outputs.
\section{Extra Samples}
\label{sec:extra-samples}

Additional graded samples, shown with the same real-image / grade-schematic pairing used in
Figs.~\ref{fig:traj1} and~\ref{fig:traj2}.

\begin{figure*}
\centering
\includegraphics[width=\textwidth]{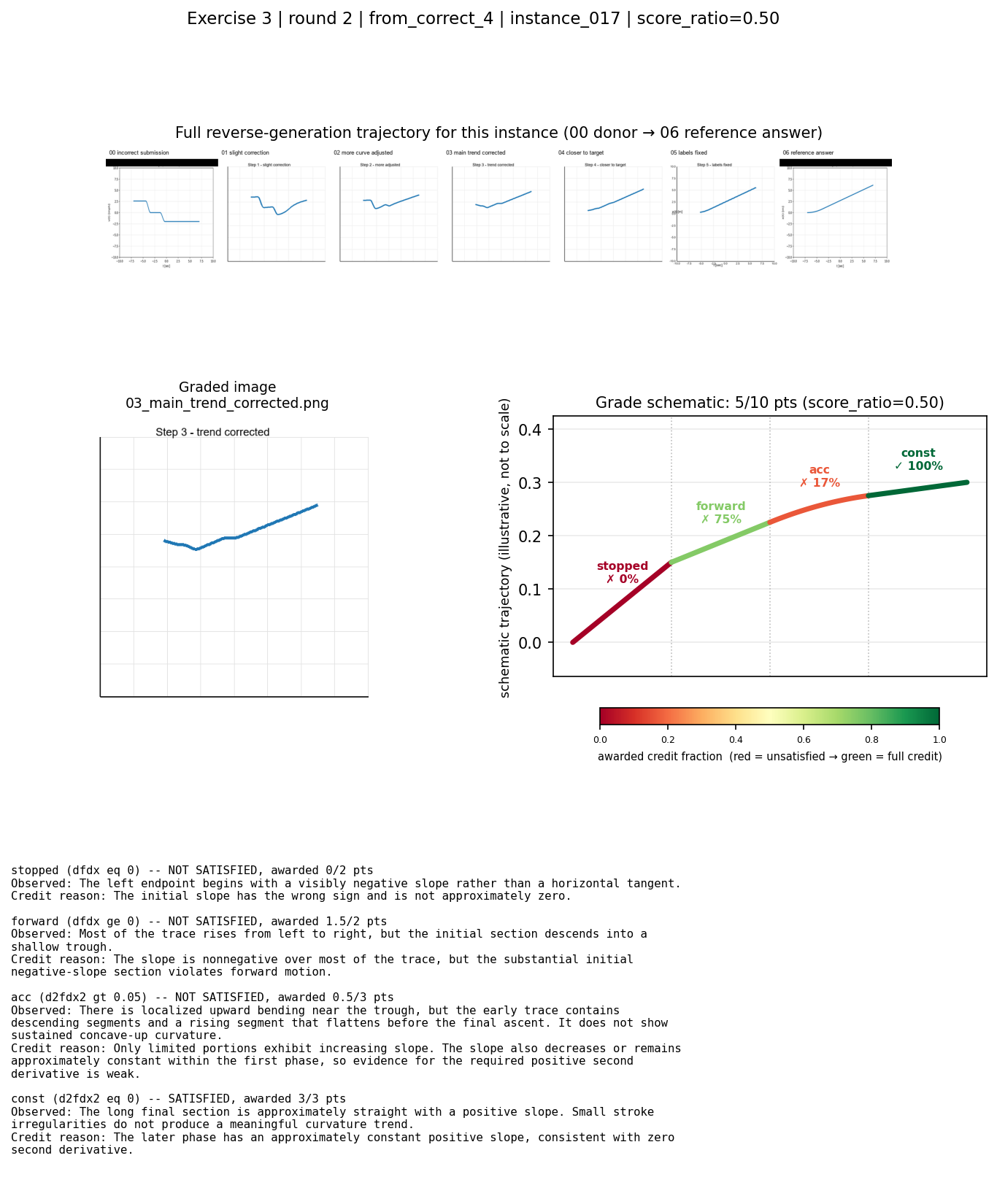}
\caption{Exercise 3, instance\_017 (source: Exercise 4), step 3, score\_ratio $=0.50$.}
\end{figure*}

\begin{figure*}
\centering
\includegraphics[width=\textwidth]{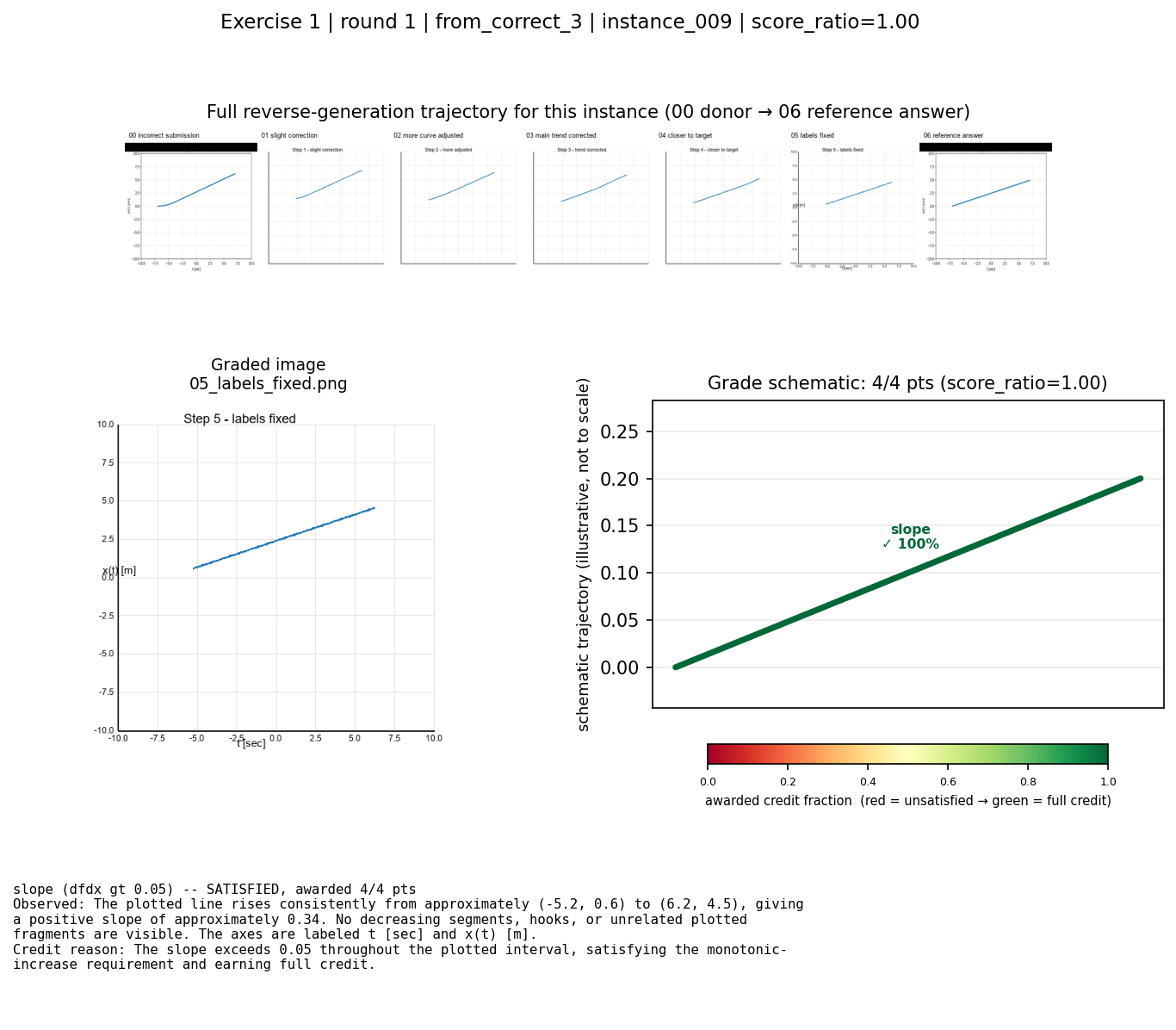}
\caption{Exercise 1, instance\_009 (source: Exercise 3), step 5, score\_ratio $=1.00$.}
\end{figure*}

\begin{figure*}
\centering
\includegraphics[width=\textwidth]{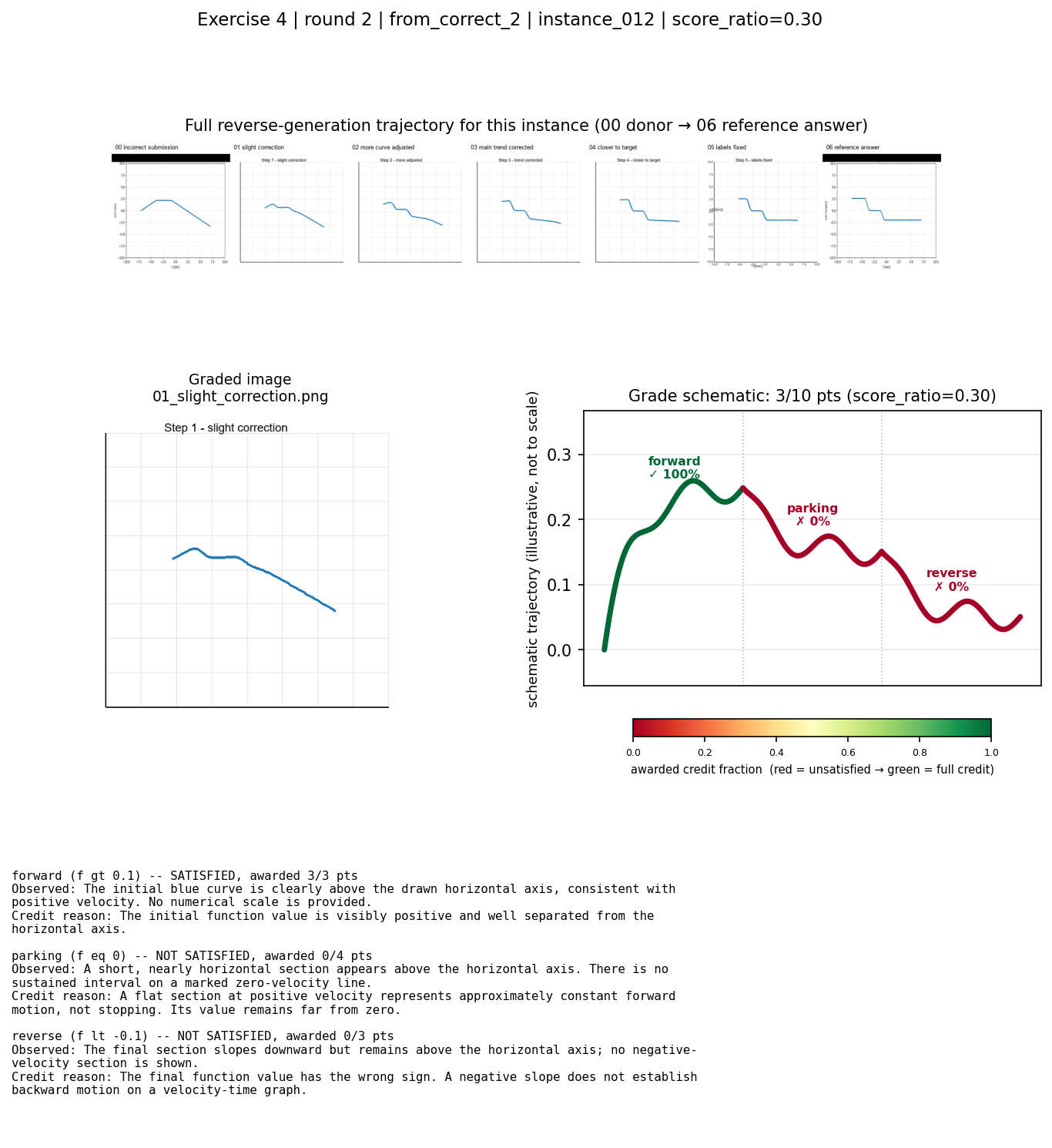}
\caption{Exercise 4, instance\_012 (source: Exercise 2), step 1, score\_ratio $=0.30$.}
\end{figure*}

\bibliography{references}
\end{document}